\documentclass[runningheads]{llncs}
\usepackage[T1]{fontenc}
\usepackage{graphicx}
\usepackage{color}
\usepackage{hyperref}

\usepackage{multirow}
\usepackage{bm}
\usepackage{algorithm}
\usepackage{algorithmic}
\usepackage{booktabs}
\usepackage{amsmath}
\usepackage{amsthm}
\usepackage{amssymb}
\usepackage{mathtools}
\usepackage{tikz}
\usepackage{xcolor}
\usetikzlibrary{arrows}
\usepackage{nicefrac} 

\usepackage{subcaption}
\usepackage{xspace}

\newcommand{\citep}{\cite}
\newcommand{\citet}{\cite}
\newcommand{\INDSTATE}[1][1]{\STATE\hspace{#1\algorithmicindent}}
\usepackage[misc]{ifsym}

\begin{document}
\title{Stock Trading Volume Prediction with Dual-Process Meta-Learning}
%
%
\author{Ruibo Chen\inst{1} \and
Wei Li\inst{2} \and
Zhiyuan Zhang\inst{1} \and
Ruihan Bao\inst{3,\textrm{(\Letter)}} \and
Keiko Harimoto\inst{3} \and
Xu Sun\inst{1,\textrm{(\Letter)}}}
\authorrunning{R. Chen et al.}

\institute{Peking University, Beijing, China \\
\email{\{ruibochen,zzy1210,xusun\}@pku.edu.cn} \and
Beijing Language and Culture University, Beijing, China\\
\email{liweitj47@blcu.edu.cn} \and
Mizuho Securities Co., Ltd\\
\email{\{ruihan.bao,keiko.harimoto\}@mizuho-sc.com} }


\toctitle{Stock Trading Volume Prediction with Dual-Process Meta-Learning}
\tocauthor{Ruibo Chen, Wei Li, Zhiyuan Zhang, Ruihan Bao, Keiko Harimoto and Xu Sun}

\maketitle              

\begin{abstract}

Volume prediction is one of the fundamental objectives in the Fintech area, which is helpful for many downstream tasks, e.g., algorithmic trading. Previous methods mostly learn a universal model for different stocks. However, this kind of practice omits the specific characteristics of individual stocks by applying the same set of parameters for different stocks. On the other hand, learning different models for each stock would face data sparsity or cold start problems for many stocks with small capitalization. To take advantage of the data scale and the various characteristics of individual stocks, we propose a dual-process meta-learning method that treats the prediction of each stock as one task under the meta-learning framework. Our method can model the common pattern behind different stocks with a meta-learner, while modeling the specific pattern for each stock across time spans with stock-dependent parameters. Furthermore, we propose to mine the pattern of each stock in the form of a latent variable which is then used for learning the parameters for the prediction module. This makes the prediction procedure aware of the data pattern. Extensive experiments on volume predictions show that our method can improve the performance of various baseline models. Further analyses testify the effectiveness of our proposed meta-learning framework. 


\end{abstract}

\section{Introduction}
Stock trading volume prediction is one of the fundamental objectives in the Fintech area, which plays a crucial role in various downstream tasks, e.g., algorithmic trading. Volume prediction aims to predict the absolute volume value or the movement trend within a certain period of time based on the historical trading-related information. 
Considering the importance of volume prediction, many researchers have been devoted to predicting the volume. Both classical machine learning models and deep learning models have been applied in volume prediction. For instance, Liu and Lai\citep{liu2017intraday} propose to predict the volume with the dynamic SVM method. Chen, Feng, and Palomar\citep{chen2016forecasting} propose to adopt a Kalman filter approach. While Libman, Haber,and Schaps\citep{libman2019volume} first propose to apply the LSTM models in volume pre-diction, which is popularly used for sequence prediction. 

Although these methods have produced practicable prediction results, they basically model different stocks with one universal set of parameters. This kind of approach omits the individual characteristics of each stock. For example, the volumes of stocks with different scales of capitalization or from different industries can follow quite different movement patterns. On the other hand, learning different sets of parameters for each stock would face severe data sparsity and cold start problems, especially for newly listed stocks. 

Based on the above observations, we propose to introduce the meta-learning framework into volume prediction. Under the proposed meta-learning framework, we propose to treat each stock as one individual task, while a meta-learner is responsible for learning the general pattern from the whole market. The meta-learner is updated according to the learning process of each task, so that its parameters can stay sensitive to individual tasks. 

Apart from the pattern variation among different stocks, we assume that the pattern of one stock from different time spans can vary too. Therefore, we propose a dual meta-learning process that makes the parameters not only sensitive to different stocks (tasks), but also sensitive to different time spans. 
To model the movement pattern of each stock at a specific period, we propose to learn a latent variable for each sampled batch from that period of time with an encoder. This latent variable is then fed into a decoder to produce the actual prediction parameters. Note that the encoder-decoder framework instead of the prediction model plays the role of meta-learner in our method. This dual meta-learning process makes the latent variable sensitive to different time periods inside given the given stock while the decoder sensitive to different stocks (tasks).

To test the effectiveness of our proposed dual meta-learning process method, we conduct experiments on the TPX500 volume prediction dataset. Extensive analyses show that our dual meta-learning process outperforms the traditional methods and neural network baselines on five-minute and ten-minute dataset. Our codes have been made public.\footnote[1]{\url{https://github.com/RayRuiboChen/DPML}}


We conclude our contributions as follows:
\begin{itemize}
    \item We propose to introduce the meta-learning framework into the volume prediction task to take advantage of both the general pattern and the individual stock patterns. In order to model the specific patterns of each stock, we apply an encoder-decoder framework, which encodes the volume variation trend into a latent variable.
    \item We propose a dual meta-learning process method to make the meta-learner sensitive to both the task-specific pattern and the time-specific pattern.
    \item Experiment results show that our proposed method can significantly improve the performance of various popularly applied baseline models.
\end{itemize}

\section{Related Work}
\subsection{Stock Market Prediction}
As deep learning techniques developed rapidly in recent years, much effort has been made in the finance area, such as stock market prediction. Existing methods are mainly based on classic models, such as Feedforward Neural Networks (FNN)\citep{chen2017double,song2019study}, Convolutional Neural Networks (CNN)\citep{sezer2018algorithmic,sim2019deep}, Recurrent Neural Networks (RNN), including Gated Recurrent Unit model (GRU)\citep{liu2019combining} and Long-Short Term Memory model (LSTM)\cite{nelson2017stock,siami2019comparative}. Liu et al.\citep{liu2019transformer} first use Capsule Network based on Transformer Encoder to predict stock movements. Ding et al.\citep{ding2020hierarchical} propose several enhancements to the basic Transformer in stock movement prediction.

\subsection{Meta-Learning}
Recent meta-learning approaches can be basically classified into three categories, metric-based, model-based and optimization-based techniques.

Metric-based methods like Siamese networks\citep{koch2015siamese} use neural networks to map the input into a feature space, and predict labels by comparing the similarity between features from support sets and query sets. Matching networks\citep{vinyals2016matching} absorbs the same idea and learns a network to map the support sets and unlabelled examples to their labels. Cosine similarity is used and they are trained in the few-shot setting. Prototypical networks\citep{snell2017prototypical} generate a prototype for each class in the feature space for comparing, increasing the robustness and reducing the time for inference. Relation networks\citep{sung2018learning} propose to use a network to work as the similarity function, which breaks the limits of pre-defined similarity metrics and exploits the task-specific information.

Model-based techniques usually use a fixed neural network at test time, and use various memory techniques to store the information from previously seen inputs or tasks. Meta Networks\citep{munkhdalai2017meta} use fast weights and slow weights to generate task-specific weights. SNAIL\citep{mishra2017simple} use the temporal convolution and attention mechanisms to improve memory capacity.

Optimization-based techniques are aimed at learning new tasks quickly with optimization methods and they mostly view meta-learning as a bi-level optimization problem. In the inner level(usually described as the inner loop), a base learner is proposed to make task-specific adjustments and the outer level(the outer loop) is concerned with performance across tasks. Model Agnostic Meta-Learning (MAML)\citep{finn2017model} uses second-order derivatives to find the most sensitive parameters in the parameter space for fast adaptation to new tasks. A large number of variations\citep{nichol2018reptile,grant2018recasting,rajeswaran2019meta,finn2019online} are proposed afterwards. Meta-SGD\citep{li2017meta} learns a learning rate vector and aims to adapt to the given task in one optimization step. 
Latent Embedding Optimization\citep{rusu2018meta} proposes an encoder-decoder architecture and optimizes in the latent embedding space under the few-shot setting.

However, previous methods mostly concentrate on classification tasks and are more suitable for few-shot learning. When presented with a larger dataset, they often cannot perform well and are computationally expensive\citep{hospedales2020meta}. Thus, we propose the dual meta-learning process, which are able to solve both classification and regression problems and can deal with the few-shot setting as well as large support set scenarios.

\begin{algorithm}
\begin{algorithmic}
\caption{Meta-Train}
\label{alg:meta-train}
\REQUIRE Stocks $S$, Encoder $e$, Decoder $d$, prediction model $f$, learning rates $\alpha,\beta, \gamma$

Initialize encoder parameters $\phi_e$, decoder parameters $\phi_d$

\textbf{for} i=1,2,... \textbf{do}

\INDSTATE For $S_i=(D^{train}_i,D^{test}_i)$, $z_i$ is the latent variable of $S_i$

\INDSTATE $\phi_{d_i}=\phi_d$

\INDSTATE \textbf{for} a few steps

\INDSTATE[2] Sample a time span $t_1$ with batch $\{x_{t_1}, y_{t_1}\}$ from $D^{train}_i$

\INDSTATE[2] $z=e(\phi_e,x_{t_1})$

\INDSTATE[2] \textbf{for} a few steps:

\INDSTATE[3] $\theta'=d(\phi_{d_i},z)$

\INDSTATE[3] $\mathcal{L}_1 = loss(f(\theta', x_{t_1}),y_{t_1})$

\INDSTATE[3] $z = z - \alpha\nabla_{z}\mathcal{L}_1	$

\INDSTATE[2] \textbf{end for}
\INDSTATE[2] $z_i = z_i + \beta(z-z_i)$
\INDSTATE[2] $\theta_i = d(\phi_{d_i},z_i)$
\INDSTATE[2] Sample another time span $t_2$ with batch $\{x_{t_2}, y_{t_2}\}$ from $D^{train}_i$
\INDSTATE[2] $\mathcal{L}_2 = loss(f(\theta_i, x_{t_2}),y_{t_2})$
\INDSTATE[2] Update $\phi_e,\phi_{d_i} $using $\mathcal{L}_2$
\INDSTATE \textbf{end for}

\INDSTATE $\phi_d=\phi_d+\gamma(\phi_{d_i}-\phi_d)$

\textbf{end for}

\end{algorithmic}
\end{algorithm}

\section{Approach}

\begin{figure}[htbp]
    \centering
    \includegraphics[height=2in,width=0.8\linewidth]{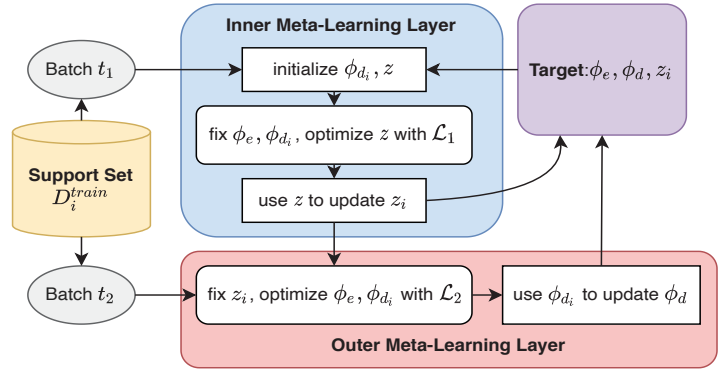}
    \caption{An overview of the optimization steps in the dual meta-learning process.}
    \label{fig:optimization}
\end{figure}

In this paper, we propose the dual meta-learning approach on top of the encoder-decoder framework. The encoder-decoder framework is responsible for extracting the patterns behind the data and learning latent variables $z$ representing the task data distribution, while the dual meta-learning process is to keep the model parameters sensitive to different stocks from different time spans. Our model first generates latent variables $z$ for each stock with the help of the encoder, which represent the characteristics of the stocks. Then we calculate the parameters $\theta$ for actual prediction models through the decoder using latent variables $z$. A dual meta-learning process with two layers is proposed to endow the encoder-decoder framework with the ability to learn the features and similarities among stocks regarding both stock level and time scale level. The inner meta-learning layer optimizes $z$ for each stock by learning different time spans while the outer layer focuses on different stocks and meta-learn through the encoder-decoder framework, making the parameters sensitive to changes, such that the model can quickly adapt to different tasks. Fig.\ref{fig:optimization} visualizes the whole optimization steps illustrated in algorithm \ref{alg:meta-train}.

\subsection{Encoder-Decoder Framework}

Instead of learning high-dimensional prediction module parameters $\theta$ directly, we propose to apply the encoder-decoder framework to learn the pattern behind stock data with latent variables $z$, which is represented in the form of low-dimensional vectors for each stock. The encoder maps the input data $x$ to $z$ in the latent space, and $z$ serves as the input of the decoder. Note that the parameters $\theta$ for the prediction module $f$ are produced by the decoder, and the final prediction result $y'$ can be computed as $y'=f(\theta,x)$.

\subsubsection{Encoder}
The encoder can effectively capture the patterns behind data and transform the knowledge into low-dimensional latent variables. It takes input data $x$ as input and generate latent variables. Given a batch of data ${x,y}$, the encoder $e$ together with its parameters $\phi_e$, we calculate the latent variable $z$ as:
\begin{equation}
z=e(\phi_e,x)\label{encoder}
\end{equation}
The latent variables contain local information and patterns for each batch with unique time spans, and they will then be generalized in the inner meta-learning layer process to produce the latent variable for the whole stock.

\subsubsection{Decoder}
The decoder is designed to output proper prediction model parameters $\theta$ based on different latent variables for different stocks. Instead of treating all stocks uniformly, the proposed decoder makes every stock attached with its own prediction model parameters, which makes the prediction module sensible to the pattern of individual stocks. The decoder works as:
\begin{equation}
\theta=d(\phi_d,z)\label{decoder}
\end{equation}
where $d$ and $\phi_d$ represent the decoder and its parameters, $z$ is the latent variable fed into the decoder.

\begin{figure}[htbp]
    \centering
    \subcaptionbox{The inner meta-learning layer}{\includegraphics[height=1in,width=0.59\linewidth]{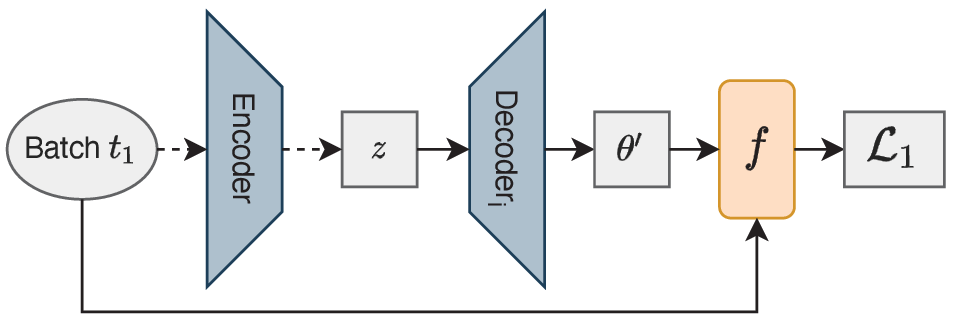}}
    \hfill
    \subcaptionbox{The outer meta-learning layer}{\includegraphics[height=1in,width=0.39\linewidth]{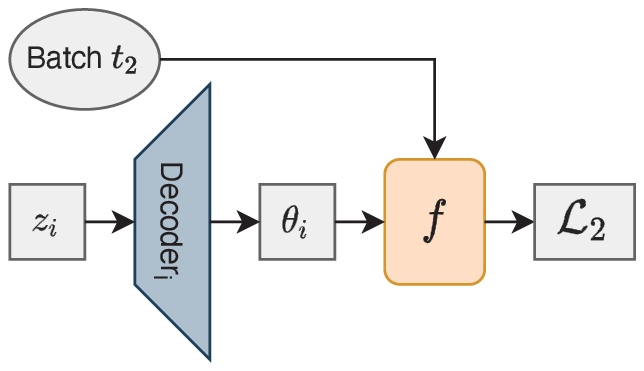}}
    \caption{A visualization of how the encoder-decoder framework interacts with the dual meta-learning process. Note that in the inner layer, the encoder is only used to initialize the latent variable $z$, linked by dashed line. In the outer layer, the encoder is not involved. The gradients of the encoder parameters $\phi_e$ can be backpropagated through $z$.}
    \label{fig:encoder-decoder with dual meta-learning}
\end{figure}

\subsection{Dual Meta-Learning Process}

Different stocks and different time spans compose two major challenges for the stock volume prediction task. Stocks are heavily influenced by the companies' actual performance, while time span features can vary according to accidental events or policies. Thus, we propose the dual meta-learning process to make our model better utilize the specific characteristics of different stocks and time periods.

Intuitively, we separate the process into two layers. The inner meta-learning layer is intended to learn the pattern behind different time spans and to make the model more precise and robust when handling new time spans in the future. The outer layer focuses on different patterns behind different stocks and makes sure the model gains sufficient global knowledge while learning individual features.

Note that our dual meta-learning process is fundamentally different from the traditional bi-level setting in optimization-based techniques, as the inner layer uses a meta-learning approach to meta-learn inside a single task, and the outer layer resembles the classic bi-level problem. A tri-level setting among instance level(time span), task level(single task) and task distribution level(all tasks) is actually proposed and processed.

As we store latent variables $z_i$ for stocks, and optimize the model in the latent space for $z_i$ and the parameter space for the encoder-decoder framework, which will be shown in the following sections, our dual meta-learning process has the feature for both model-based techniques and optimization-based techniques. The detailed architectures of the dual-process meta-learning are shown in Fig.\ref{fig:encoder-decoder with dual meta-learning}.

\subsubsection{Inner Meta-Learning Layer}
The inner meta-learning layer mainly functions inside different time spans in one stock. To generate a stock latent variable that is sensitive to time spans, we do not use the whole training data in the support set, which can be large, time consuming and can omit the information for small time scales. Instead, we sample a batch of data $\{x_{t_1},y_{t_1}\}$ which are continuous in time and represent the stock pattern during the given time span $t_1$.

The inner layer works by incorporating the characteristics of latent variables $z$ for each batch into the stock latent variable $z_i$. For each $z$ initialized by the encoder, we first optimize it by using inner meta-train loss $\mathcal{L}_1$, which is computed as:

\begin{equation}
\theta'=d(\phi_d,z)
\end{equation}
\begin{equation}
\mathcal{L}_1=loss(f(\theta',x_{t_1}),y_{t_1})
\end{equation}
Note that all other parameters like $\phi_e$, $\phi_d$ are kept fixed in the inner layer's meta-learning procedure.

After a few steps, we add the underlying information for the certain time span in optimized $z$ into the latent variable for the i-th stock $z_i$ by:
\begin{equation}
z_i=z_i+\beta(z-z_i)
\end{equation}

\subsubsection{Outer Meta-Learning Layer} In contrast to the inner layer meta-learning process, the outer layer is designed to learn the pattern behind different stocks. The encoder-decoder framework and distinctive stock latent variables introduce a large amount of uniqueness, and the outer meta-learning layer secures enough global knowledge by learning and generalizing comprehensive patterns between stocks, allowing quick adaptation to new tasks.

We keep global encoder parameters $\phi_e$ and decoder parameters $\phi_d$ across different stocks in the outer meta-learning layer. To obtain the similarities in different stocks, in every epoch we create unique decoder parameters $\phi_{d_i}$ for the i-th stock and initialize it with $\phi_d$. Similar to the inner layer, the outer layer also uses optimized $\phi_{d_i}$ to carry stock-specific knowledge.

Note that in the outer meta-learning layer, we sample another batch of data $\{x_{t_2},y_{t_2}\}$ from a different time span $t_2$. We do not directly use $\{x_{t_1},y_{t_1}\}$ sampled in the inner layer in order to avoid overfitting on the same time span and enhance the generalization ability of the model. Given $\phi_{d_i}$ and optimized $z_i$, the outer meta-train loss $\mathcal{L}_2$ is computed as:
\begin{equation}
\theta_i=d(\phi_{d_i},z_i)
\end{equation}
\begin{equation}
\mathcal{L}_2=loss(f(\theta_i,x_{t_2}),y_{t_2})
\end{equation}
In the outer meta-learning layer, we only update $\phi_e$, $\phi_{d_i}$ using $\mathcal{L}_2$ and keep $z_i$ fixed. The alternate optimization separated in two layers similar to Generative Adversarial Networks helps layers to reach local optima in each step and finally move to global optima during the meta-training procedure.

After $\phi_{d_i}$ is optimized, we tune $\phi_d$ towards $\phi_{d_i}$ in the parameter space by:
\begin{equation}
\phi_d=\phi_d+\gamma(\phi_{d_i}-\phi_d)
\end{equation}
In this process, $\phi_d$ meta-learn the differences between different stocks with the help of first-order gradients and become sensitive in the parameter space, minimizing the expected loss across task distribution as Nichol et al.\citep{nichol2018reptile} discussed.

\subsection{Inference}

The meta-learning setup consists of meta-training, meta-development and meta-test stages. Tasks for meta-development and meta-test are not seen during the meta-train stage, thus evaluating the generalization ability of the trained model. Considering the tri-level setting proposed in our dual meta-learning process, the dataset segmentation can be done from the task level and instance level, and different inference algorithms are proposed as follows.

Segmentation in the task level coincides with traditional settings, and in the stock market prediction area,  we can simply view different stocks as different tasks. During meta-training, only part of the stocks are available, and the meta-test stage focuses on results on unseen stocks. In this time, inner layer must first be applied to acquire the latent variable $z_i$ for the new task, which can be efficiently initialized by using the mean of latent variables of meta-training tasks.

However, in application, the stock market prediction problems are mostly time series analysis problems, where all stocks are available, but time spans are restricted. We propose instance level dataset segmentation for this kind of data, that all stocks are available but time spans are divided for meta-train, meta-evaluate and meta-test in chronological order. This is more suitable in real work application and we are more concerned with the performance on the unknown, future time spans.

The inference algorithm is given in Algorithm \ref{alg:inference}. The meta-train process provides a proper representation for each stock as different latent variables and globally effective parameters for encoder and decoder. As $\phi_d$ are meta-learned and sensitive to changes in the parameter space, we optimize the $\phi_d$ using the support set for the corresponding stock for a few steps to make it quickly adapt to the given task. Then we use the prediction model parameters produced by the tuned decoder to evaluate and get the final prediction result. This process is similar to meta-learning techniques like MAML and Meta-SGD.

\begin{algorithm}
\begin{algorithmic}
\caption{Inference}
\label{alg:inference}
\REQUIRE Stocks $S$, Decoder $d$, model $f$

For $S_i=(D^{train}_i,D^{test}_i)$

$\phi_{d_i}=\phi_d$

\textbf{for} a few steps

\INDSTATE Sample a time span $t$ with batch $\{x_t, y_t\}$ from $D^{train}_i$

\INDSTATE $\theta'=d(\phi_{d_i},z_i)$

\INDSTATE $\mathcal{L}_t = loss(f(\theta', x_t),y_t)$

\INDSTATE Update $\phi_{d_i}$ using $\mathcal{L}_t$

\textbf{end for}

$\theta_i=d(\phi_{d_i},z_i)$ 

Compute $\mathcal{L}_{test}=loss(f(\theta_i,x),y)$ for ${x, y}$ in $D^{test}_i$

\end{algorithmic}
\end{algorithm}

\subsection{Model Agnostic}
An important feature of the encoder-decoder framework is that it can be easily applied to any models. For example, we can replace the last fully-connected (FC) layer with the encoder-decoder framework, where input data $x$ are the input vectors for the original last FC layer. In this situation, the given model like LSTM or Transformer can be viewed as a feature extractor. The feature can then be fed into the encoder-decoder framework to be processed. This makes our approach model-agnostic, which means that existing models can leverage our dual meta-learning process to improve performance. If a feature extractor network $F$ is used, we first pre-train the feature extractor on the meta-training dataset. Then the input batch can be presented as $\{F(x_t),y_t\}$ given time span $t$. The feature extractor can be optimized in the outer layer using $\mathcal{L}_2$ during meta-training stage.

\section{Experiment}

\subsection{Tasks and Datasets}

\subsubsection{Dataset and Data Preprocessing}
In this paper, we adopt five-minute and ten-minute intra-day volume prediction dataset. The two datasets are extracted from the Topix500 dataset with volumes and open, close, high, low prices. The input data consists of log volumes and prices of the previous 12 time slots(in the same day) and the same time slots in the previous 20 trading days. We dropped the data instances which have missing volumes or prices. The target of our prediction task is to regress the log volume.

Our data were collected between \textit{2017} and \textit{2018}. We choose the proposed instance level data segmentation to simulate the application scene. We adopt the data of \textit{2017} for meta-training set and meta-development set, and the data of\textit{Jan.2018} and \textit{Feb.2018} as the test set. The training set and development set are split by time. The statistics of the two datasets are shown in Table \ref{tab:dataset}.

\begin{table}[htbp]
	\centering
	\caption{Statistic information on the two datasets}
	\label{tab:dataset}
	\begin{tabular}{l|ccc|ccc}
		\toprule  
		\textbf{Dataset}&\multicolumn{3}{c}{\textbf{Five-minute}}&\multicolumn{3}{c}{\textbf{Ten-minute}}\\
		\midrule
		\textbf{Split}&Meta-Train&Meta-Dev&Meta-Test&Meta-Train&Meta-Dev&Meta-Test\\
		\midrule
		\textbf{Samples}&106139&35359&27189&318383&81562&76418\\
		\bottomrule
	\end{tabular}
\end{table}

\subsubsection{Evaluation Metrics}
We adopt three evaluation metrics for our volume prediction task: mean squared error(MSE), mean absolute error(MAE) and accuracy(ACC). Given input data pair $\{x,y\}$, prediction result $\hat{y}=f(\theta,x)$, the three metrics are defined as: $MSE=\mathbb{E}_{(x,y) \sim \mathcal{D}}(\hat{y}-y)^2$, $MAE=\mathbb{E}_{(x,y) \sim \mathcal{D}}|\hat{y}-y|$, $ACC=\mathbb{P}_{(x,y) \sim \mathcal{D}}((\hat{y}-v_{last})(y-v_{last})>0)$.


Here $v_{last}$ represents the volume of the last time slot and ACC is the accuracy of whether the predicted volumes vary in accordance with the ground truth compared with the last time slot.

\subsection{Baselines}

\subsubsection{Traditional Methods}
\begin{itemize}

\item \textbf{Naive forecasting.} In our experiment, the naive forecasting algorithm uses volumes of last time slot or the same slot in yesterday.

\item \textbf{Simple moving average(SMA).} The simple moving average algorithm calculates the naive average value. In our experiment, we adopt the 12-slot average, 20-day average, and 12-slot and 20-day average.

\item \textbf{Exponential moving average(EMA).} Given a series of data $\{x_1, x_2,...\}$, the EMA series $y_n(y_1=x_1)$ are computed by $y_n=\frac{2x_n+(n-1)y_{n-1}}{n+1}$. In our experiments, we tried 20-day EMA and 12-slot EMA.
\end{itemize}

\subsubsection{Linear} Given input data $x$ and model parameters $\theta$=$(w,b)$, the linear model is formulated as $f(\theta,x)=w^Tx+b$. We use the concatenation of 12-slot and 20-day history as $x$ in our experiments.

\subsubsection{LSTM} Following the widely use of LSTM\citep{nelson2017stock,siami2019comparative} in stock market prediction task, we implement two one-layer LSTM models for previous 12-slot and 20-day history respectively. First, we project the input data to a feature space using an FC layer. Then the features are fed into the LSTM models, followed by an attentive pooling layer. Then another FC layer is used to get the prediction result.

\subsubsection{Transformers} We also implement a six-layer Transformer Encoder\cite{vaswani2017attention} model as a baseline. The input data consists of a special \textit{[CLS]} token and the concatenation of the 12-slot and 20-day data. The Positional Encoding is enabled. The prediction result is computed by using the output vector of \textit{[CLS]} token to feed into a FC layer.

\begin{table*}[htbp]
	\centering
	\caption{Experimental results}
	\label{tab:res}
	\begin{tabular}{l|ccc|ccc}
		\toprule
		\textbf{Dataset}&\multicolumn{3}{c}{\textbf{Five-Minute}}&\multicolumn{3}{c}{\textbf{Ten-Minute}}\\
		\toprule
		\textbf{Model}&\textbf{MSE$\downarrow$}&\textbf{MAE$\downarrow$}&\textbf{ACC$\uparrow$}&\textbf{MSE$\downarrow$}&\textbf{MAE$\downarrow$}&\textbf{ACC$\uparrow$}\\ 
		\midrule
		Yesterday
		&1.203&0.797&0.665&0.517&0.532&0.719\\
		20-day Average
		&0.698&0.607&0.709&0.433&0.503&0.720\\
		20-day EMA
		&0.689&0.600&0.713&0.427&0.498&0.727\\
		Last Time Slot
		&1.118&0.742&0.500&0.653&0.602&0.500\\
		12-slot Average
		&0.982&0.710&0.630&0.975&0.782&0.445\\
		12-slot EMA
		&0.888&0.668&0.642&0.846&0.718&0.457\\
		20-day and 12-slot Average
		&0.689&0.581&0.713&0.377&0.469&0.698\\
		\midrule
		Linear
		&$0.694$&$0.638$&$0.681$&$0.303$&$0.419$&$0.740$\\
		Linear+ours
		&$0.623$&$0.585$&$0.710$&$0.266$&$0.381$&$0.760$\\
		LSTM
		&$0.623$&$0.583$&$0.706$&$0.272$&$0.391$&$0.745$\\
		LSTM+ours
		&$\textbf{0.586}$&$0.556$&$\textbf{0.724}$&$\textbf{0.252}$&$\textbf{0.370}$&$\textbf{0.765}$\\
		Transformer
		&$0.611$&$0.573$&$0.711$&$0.270$&$0.389$&$0.748$\\
		Transformer+ours
		&$0.589$&$\textbf{0.555}$&$\textbf{0.724}$&$0.255$&$0.372$&$0.764$\\
		\bottomrule
	\end{tabular}
\end{table*}

\subsection{Settings and Hyperparameters}
We repeat every experiment for 5 times and report the result on the meta-test dataset on the checkpoint with the lowest meta-development MSE loss. For hyperparameters in algorithm \ref{alg:meta-train}, we set $\alpha$=1e-4, $\beta$=1e-4, $\gamma$=1 and the stock latent variables are initialized to zeros. We adopt the SGD optimizer to optimize encoder parameters $\phi_e$ and decoder parameters $\phi_{d_i}$ with the learning rate set to 1e-5. For encoder $e$, decoder $d$, we adopt Multilayer Perceptron(MLP) with 3 layers. For prediction model $f$, we use a linear model. The loss function we used in algorithm \ref{alg:meta-train} and algorithm \ref{alg:inference} is MSE loss. For baseline models and pre-train stage for feature extractors, we adopt the Adam optimizer with the learning rate initialized to 1e-4. The batch size we used is 32. In meta-development and meta-test stages, we only conduct 10 steps in tuning $\phi_{d_i}$ and we use the SGD optimizer with the learning rate set to 1e-6.

\subsection{Experimental Results}
After selecting the best hyperparameter configurations based on the results on the meta-development set, the experimental results on meta-test set are shown in Table \ref{tab:res}. As the result illustrated, our methods successfully improves the performance on three neural network baselines in both five-minute and ten-minute tasks. They also remarkably outperform the traditional baseline results.


\begin{figure}[!h]
    \centering
    \subcaptionbox{Small market capitalization stocks on five-minute dataset}{\includegraphics[height=1.2in,width=0.45\linewidth]{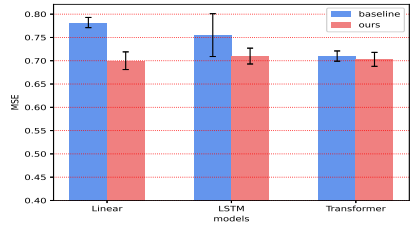}}
    \hfill
    \subcaptionbox{Large market capitalization stocks on five-minute dataset}{\includegraphics[height=1.2in,width=0.45\linewidth]{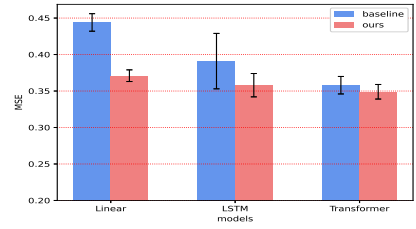}}
    
    \subcaptionbox{Small market capitalization stocks on ten-minute dataset}{\includegraphics[height=1.2in,width=0.45\linewidth]{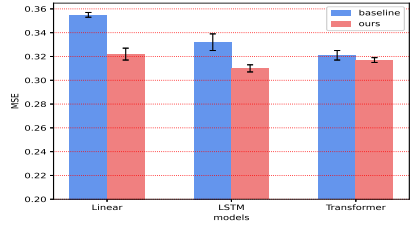}}
    \hfill
    \subcaptionbox{Large market capitalization stocks on ten-minute dataset}{\includegraphics[height=1.2in,width=0.45\linewidth]{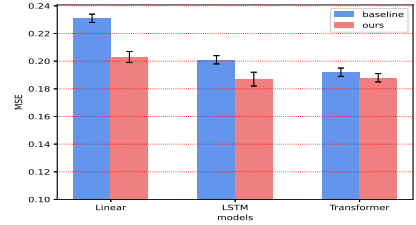}}
    
    \caption{MSE loss on simulation of newly listed stocks. On both dataset our dual meta-learning process enhance the performance of three baselines. Note that the loss for stocks with small market capitalization is significantly higher than those with large market capitalization, indicating that they are more sensitive and difficult to predict.}
    \label{fig:reduce_data}
\end{figure}

\section{Analysis}
\subsection{Effectiveness of Meta-learning}

In table \ref{tab:meta-learning}, we show the experiment results on whether treating different stocks as different tasks in the meta-learning framework(w/o tasks). We can see that modeling different stocks with stock-specific parameters yield large gain on all the metrics in both five-minute and ten-minute datasets. This testifies the assumption that different stocks vary in the volume variation trend. Therefore, modeling stocks with stock-specific parameters is necessary.

To test whether our meta-learning method can improve the model performance on newly listed stocks, where the number of historical volume data is small, we conduct experiments to simulate those cases on five-minute and ten-minute datasets. We randomly sample 50 stocks with relatively large market capitalization or small market capitalization and only keep their last 10\% data in chronological order. We report the meta-test MSE loss with lowest meta-development MSE loss.

From the results shown in figure \ref{fig:reduce_data} , we can see that applying the meta-learning framework can indeed enhance the model performance on few-shot cases especially for less effective baseline models, linear and LSTM. Whereas the performance of the Transformer baseline also improves. Furthermore, the gap between stocks with large and small market capitalization proves that different patterns exist in different stocks, which can be seized by the dual meta-learning process.

\begin{table*}[htbp]
	\centering
	\caption{Ablation Study: Effectiveness of Meta-learning}
	\label{tab:meta-learning}
	\begin{tabular}{l|ccc|ccc}
		\toprule  
		\textbf{Dataset}&\multicolumn{3}{c}{\textbf{Five-Minute}}&\multicolumn{3}{c}{\textbf{Ten-Minute}}\\
		\toprule
		\textbf{Model}&\textbf{MSE$\downarrow$}&\textbf{MAE$\downarrow$}&\textbf{ACC$\uparrow$}&\textbf{MSE$\downarrow$}&\textbf{MAE$\downarrow$}&\textbf{ACC$\uparrow$}\\ 
		\midrule  
		Linear
		&$0.694$&$0.638$&$0.681$&$0.303$&$0.419$&$0.740$\\
		+our approach
		&$\textbf{0.623}$&$\textbf{0.585}$&$\textbf{0.710}$&$\textbf{0.266}$&$\textbf{0.381}$&$\textbf{0.760}$\\
		w/o tasks
		&$0.745$&$0.669$&$0.669$&$0.315$&$0.430$&$0.731$\\
		\midrule
		LSTM
		&$0.623$&$0.583$&$0.706$&$0.272$&$0.391$&$0.745$\\
		+our approach
		&$\textbf{0.586}$&$\textbf{0.556}$&$\textbf{0.724}$&$\textbf{0.252}$&$\textbf{0.370}$&$\textbf{0.765}$\\
		w/o tasks
		&$0.635$&$0.592$&$0.701$&$0.271$&$0.391$&$0.745$\\

		\midrule
		Transformer
		&$0.611$&$0.573$&$0.711$&$0.270$&$0.389$&$0.748$\\
		+our approach
		&$\textbf{0.589}$&$\textbf{0.555}$&$\textbf{0.724}$&$\textbf{0.255}$&$\textbf{0.372}$&$\textbf{0.764}$\\
		w/o tasks
		&$0.608$&$0.572$&$0.713$&$0.269$&$0.389$&$0.748$\\
		\bottomrule
	\end{tabular}
\end{table*}

\subsection{Effectiveness of Encoder-Decoder Framework }

\begin{table*}[htbp]
	\centering
	\caption{Ablation Study: Effectiveness of Encoder-Decoder Framework}
	\label{tab:encoder_decoder}
	\begin{tabular}{l|ccc|ccc}
		\toprule  
		\textbf{Dataset}&\multicolumn{3}{c}{\textbf{Five-Minute}}&\multicolumn{3}{c}{\textbf{Ten-Minute}}\\
		\toprule
		\textbf{Model}&\textbf{MSE$\downarrow$}&\textbf{MAE$\downarrow$}&\textbf{ACC$\uparrow$}&\textbf{MSE$\downarrow$}&\textbf{MAE$\downarrow$}&\textbf{ACC$\uparrow$}\\ 
		\midrule  
		Linear
		&$0.694$&$0.638$&$0.681$&$0.303$&$0.419$&$0.740$\\
		+our approach
		&$\textbf{0.623}$&$\textbf{0.585}$&$\textbf{0.710}$&$\textbf{0.266}$&$\textbf{0.381}$&$\textbf{0.760}$\\
		w/o encoder
		&$0.700$&$0.642$&$0.681$&$0.288$&$0.405$&$0.748$\\
		w/o encoder,latent variables
		&$0.661$&$0.610$&$0.697$&$0.272$&$0.387$&$0.758$\\
		w/o encoder,decoder
		&$0.718$&$0.656$&$0.673$&$0.292$&$0.406$&$0.747$\\
		\midrule
		LSTM
		&$0.623$&$0.583$&$0.706$&$0.272$&$0.391$&$0.745$\\
		+our approach
		&$\textbf{0.586}$&$\textbf{0.556}$&$\textbf{0.724}$&$\textbf{0.252}$&$\textbf{0.370}$&$\textbf{0.765}$\\
		w/o encoder
		&$0.630$&$0.587$&$0.704$&$0.265$&$0.383$&$0.755$\\
		w/o encoder,latent variables
		&$0.614$&$0.576$&$0.711$&$0.270$&$0.384$&$0.756$\\
		w/o encoder,decoder
		&$0.620$&$0.581$&$0.722$&$0.266$&$0.383$&$0.757$\\
		\midrule
		Transformer
		&$0.611$&$0.573$&$0.711$&$0.270$&$0.389$&$0.748$\\
		+our approach
		&$\textbf{0.589}$&$\textbf{0.555}$&$\textbf{0.724}$&$0.255$&$0.372$&$0.764$\\
		w/o encoder
		&$0.610$&$0.573$&$0.712$&$0.260$&$0.378$&$0.760$\\
		w/o encoder,latent variables
		&$0.599$&$0.564$&$0.719$&$\textbf{0.252}$&$\textbf{0.370}$&$\textbf{0.768}$\\
		w/o encoder,decoder
		&$0.603$&$0.566$&$0.718$&$0.265$&$0.381$&$0.760$\\
		\bottomrule
	\end{tabular}
\end{table*}

To test whether the encoder can extract useful information for volume prediction, we remove the encoder, where latent variables are initialized by the input features. From the results in table \ref{tab:encoder_decoder}, we can see that without the encoder module, all the metrics decline, which shows the effectiveness of our proposed encoder structure.

We further remove the design for latent variables in equation \ref{decoder} , where parameters $\theta$ are generated by the input data using the decoder directly. The performance drop indicates that latent variables $z$ are more informative in the latent space, which may work by denoising the raw input and extracting important features. On ten-minute dataset the Transformer model performance gets slightly better. It may be caused by data homogeneity in ten-minute dataset and the Transformer model may partially learn the role of the encoder.

To examine whether the design for producing prediction parameters based on the latent variables can help volume prediction, we further remove the decoder in addition to the encoder. In this case, only first-order gradients for the parameters of prediction model $f$ are exploited, degenerate into simple Reptile. In this case, performance deteriorates greatly, proving the effectiveness of the encoder-decoder framework.

\subsection{Analyzing Dual Meta-Learning Process}



In table \ref{tab:dual process}, we analyze the effectiveness of the dual meta-learning process. We first remove the inner meta-learning layer(w/o inner meta-learning) by generating the latent variable $z_i$ with the entire $D^{train}_i$ from stock $S_i$. Results show that it reduces the performance on both five-minute and ten-minute dataset. It proves that different time spans have distinct patterns and the inner meta-learning process successfully captures and exploits the features behind a small time scale.

For the outer meta-learning layer, if it is fully removed, the situation can be viewed as there is only one single task and results collapse as we have discussed before. We further probe the influence of unique decoders(w/o unique decoder). Recall that in outer meta-learning layer, we implement stock-specific decoder parameter $\phi_{d_i}$ in meta-training stage. If we replace it with a universal decoder parameter, it can be seen that on five-minute dataset, all the metrics degrade, showing that on this time scale stock-specific information can be valuable and unique decoders are influential. But on ten-minute dataset, the accuracy metric and the more effective Transformer model showed a marginal improvement in performance, which may be caused by less noise and uncertainty in the data.

\begin{table*}[htbp]
	\centering
	\caption{Ablation study: Analyzing Dual Meta-Learning Process}
	\label{tab:dual process}
	\begin{tabular}{l|ccc|ccc}
		\toprule  
		\textbf{Dataset}&\multicolumn{3}{c}{\textbf{Five-Minute}}&\multicolumn{3}{c}{\textbf{Ten-Minute}}\\
		\toprule
		\textbf{Model}&\textbf{MSE$\downarrow$}&\textbf{MAE$\downarrow$}&\textbf{ACC$\uparrow$}&\textbf{MSE$\downarrow$}&\textbf{MAE$\downarrow$}&\textbf{ACC$\uparrow$}\\ 
		\midrule  
		Linear
		&$0.694$&$0.638$&$0.681$&$0.303$&$0.419$&$0.740$\\
		+our approach
		&$\textbf{0.623}$&$\textbf{0.585}$&$\textbf{0.710}$&$\textbf{0.266}$&$\textbf{0.381}$&$0.760$\\
		w/o inner meta-learning
		&$0.664$&$0.599$&$0.705$&$0.284$&$0.398$&$0.751$\\
		w/o unique decoder
		&$0.644$&$0.599$&$0.705$&$\textbf{0.266}$&$0.382$&$\textbf{0.763}$\\
		\midrule
		LSTM
		&$0.623$&$0.583$&$0.706$&$0.272$&$0.391$&$0.745$\\
		+our approach
		&$\textbf{0.586}$&$\textbf{0.556}$&$\textbf{0.724}$&$\textbf{0.252}$&$\textbf{0.370}$&$0.765$\\
		w/o inner meta-learning
		&$0.588$&$0.557$&$0.723$&$0.253$&$0.371$&$0.764$\\
		w/o unique decoder
		&$0.594$&$0.561$&$0.720$&$0.256$&$0.372$&$\textbf{0.766}$\\
		
		\midrule
		Transformer
		&$0.611$&$0.573$&$0.711$&$0.270$&$0.389$&$0.748$\\
		+our approach
		&$\textbf{0.589}$&$\textbf{0.555}$&$\textbf{0.724}$&$0.255$&$0.372$&$0.764$\\
		w/o inner meta-learning
		&$0.664$&$0.611$&$0.699$&$0.256$&$0.374$&$0.761$\\
		w/o unique decoder
		&$0.597$&$0.561$&$0.721$&$\textbf{0.253}$&$\textbf{0.370}$&$\textbf{0.769}$\\
		\bottomrule
	\end{tabular}
\end{table*}

\section{Conclusion}
In this work, we propose the dual meta-learning process for stock trading volume prediction, which are model agnostic and can be implemented on given models without a meta-learning procedure to improve performance. We use the inner meta-learning layer to mine the pattern behind different time spans and learn a stock-specific latent variable. The outer meta-learning layer gains generalization ability across stock (task) distributions. The dual meta-learning process successfully models the characteristics of stock data and outperforms various baselines. Extensive analyses further show the effectiveness of each component of the dual meta-learning process.

\subsubsection{Acknowledgements}
We thank all the anonymous reviewers for their valuable suggestions. This work is supported by Mizuho Securities Co., Ltd. We sincerely thank Mizuho Securities for the domain expert suggestions and the experiment dataset. Ruihan Bao and Xu Sun are the corresponding authors.


%
%
%

\begin{thebibliography}{99}
\bibitem{chen2017double}
Chen, H., Xiao, K., Sun, J., Wu, S.: A double-layer neural network framework for high-frequency forecasting. ACM Transactions on Management Information
Systems (TMIS) 7(4), 1–17 (2017)

\bibitem{chen2016forecasting}
Chen, R., Feng, Y., Palomar, D.: Forecasting intraday trading volume: a kalman filter approach. Available at SSRN 3101695 (2016)

\bibitem{ding2020hierarchical}
Ding, Q., Wu, S., Sun, H., Guo, J., Guo, J.: Hierarchical multi-scale gaussian transformer for stock movement prediction. In: IJCAI. pp. 4640–4646 (2020)

\bibitem{finn2017model}
Finn, C., Abbeel, P., Levine, S.: Model-agnostic meta-learning for fast adaptation of deep networks. In: International conference on machine learning. pp. 1126–1135. PMLR (2017)

\bibitem{finn2019online}
Finn, C., Rajeswaran, A., Kakade, S., Levine, S.: Online meta-learning. In: International Conference on Machine Learning. pp. 1920–1930. PMLR (2019)

\bibitem{grant2018recasting}
Grant, E., Finn, C., Levine, S., Darrell, T., Griffiths, T.: Recasting gradient-based meta-learning as hierarchical bayes. arXiv preprint arXiv:1801.08930 (2018)

\bibitem{hospedales2020meta}
Hospedales, T., Antoniou, A., Micaelli, P., Storkey, A.: Meta-learning in   neural networks: A survey. arXiv preprint arXiv:2004.05439 (2020)

\bibitem{koch2015siamese}
Koch, G., Zemel, R., Salakhutdinov, R., et al.: Siamese neural networks for one-shot image recognition. In: ICML deep learning workshop. vol. 2, p. 0. Lille (2015)

\bibitem{li2017meta}
Li, Z., Zhou, F., Chen, F., Li, H.: Meta-sgd: Learning to learn quickly for few-shot learning. arXiv preprint arXiv:1707.09835 (2017)

\bibitem{libman2019volume}
Libman, D., Haber, S., Schaps, M.: Volume prediction with neural networks. Frontiers in Artificial Intelligence p. 21 (2019)

\bibitem{liu2019transformer}
Liu, J., Lin, H., Liu, X., Xu, B., Ren, Y., Diao, Y., Yang, L.: Transformer-based capsule network for stock movement prediction. In: Proceedings of the First Workshop on Financial Technology and Natural Language Processing. pp. 66–73 (2019)

\bibitem{liu2019combining}
Liu, J., Lu, Z., Du, W.: Combining enterprise knowledge graph and news sentiment analysis for stock price prediction. In: Proceedings of the 52nd Hawaii International Conference on System Sciences (2019)

\bibitem{liu2017intraday}
Liu, X., Lai, K.K.: Intraday volume percentages forecasting using a dynamic svm-based approach. Journal of Systems Science and Complexity 30(2), 421–433 (2017)

\bibitem{mishra2017simple}
Mishra, N., Rohaninejad, M.,Chen, X., Abbeel, P.: A simple neural attentive
meta-learner. arXiv preprint arXiv:1707.03141 (2017)

\bibitem{munkhdalai2017meta}
Munkhdalai, T., Yu, H.: Meta networks. In: International Conference on Machine Learning. pp. 2554–2563. PMLR (2017)

\bibitem{nelson2017stock}
Nelson, D.M., Pereira, A.C., De Oliveira, R.A.: Stock market's price movement prediction with lstm neural networks. In: 2017 International joint conference on neural networks (IJCNN). pp. 1419–1426. IEEE (2017)

\bibitem{nichol2018reptile}
Nichol, A., Schulman, J.: Reptile: a scalable metalearning algorithm. arXiv preprint arXiv:1803.02999 2(3), 4 (2018)

\bibitem{rajeswaran2019meta}
Rajeswaran, A., Finn, C., Kakade, S.M., Levine, S.: Meta-learning with implicit gradients. Advances in neural information processing systems 32 (2019)

\bibitem{rusu2018meta}
Rusu, A.A., Rao, D., Sygnowski, J., Vinyals, O., Pascanu, R., Osindero, S.,
Hadsell, R.: Meta-learning with latent embedding optimization. arXiv preprint arXiv:1807.05960 (2018)

\bibitem{sezer2018algorithmic}
Sezer, O.B., Ozbayoglu, A.M.: Algorithmic financial trading with deep convolutional neural networks: Time series to image conversion approach. Applied Soft Computing 70, 525–538 (2018)

\bibitem{siami2019comparative}
Siami-Namini, S., Tavakoli, N., Namin, A.S.: A comparative analysis of forecasting financial time series using arima, lstm, and bilstm. arXiv preprint arXiv:1911.09512 (2019)

\bibitem{sim2019deep}
Sim, H.S., Kim, H.I., Ahn, J.J.: Is deep learning for image recognition applicable to stock market prediction? Complexity 2019 (2019)

\bibitem{snell2017prototypical}
Snell, J., Swersky, K., Zemel, R.: Prototypical networks for few-shot learning. Advances in neural information processing systems 30 (2017)

\bibitem{song2019study}
Song, Y., Lee, J.W., Lee, J.: A study on novel filtering and relationship between input-features and target-vectors in a deep learning model for stock price prediction. Applied Intelligence 49(3), 897–911 (2019)

\bibitem{sung2018learning}
Sung, F., Yang, Y., Zhang, L., Xiang, T., Torr, P.H., Hospedales, T.M.: Learning to compare: Relation network for few-shot learning. In: Proceedings of the IEEE conference on computer vision and pattern recognition. pp. 1199–1208 (2018)

\bibitem{vaswani2017attention}
Vaswani, A., Shazeer, N., Parmar, N., Uszkoreit, J., Jones, L., Gomez, A.N., Kaiser, L., Polosukhin, I.: Attention is all you need. Advances in neural information processing systems 30 (2017)

\bibitem{vinyals2016matching}
Vinyals, O., Blundell, C., Lillicrap, T., Wierstra, D., et al.: Matching networks for one shot learning. Advances in neural information processing systems 29 (2016)



\end{thebibliography}
%

\end{document}